\documentclass[%
aip,
amsmath,amssymb,
preprint,%
]{revtex4-1}

\usepackage{graphicx}% Include figure files
\usepackage{dcolumn}% Align table columns on decimal point
\usepackage{bm}% bold math
\usepackage[utf8]{inputenc}
\usepackage[T1]{fontenc}
\usepackage{mathptmx}
\usepackage{etoolbox}

\usepackage{siunitx}

\makeatletter
\def\@email#1#2{%
	\endgroup
	\patchcmd{\titleblock@produce}
	{\frontmatter@RRAPformat}
	{\frontmatter@RRAPformat{\produce@RRAP{*#1\href{mailto:#2}{#2}}}\frontmatter@RRAPformat}
	{}{}
}%
\makeatother
\begin{document}
%	\preprint{AIP/123-QED}

	\title[The impact of IFE on AOHDS]{Exploring the impact of the inverse Faraday-effect on all-optical helicity-dependent magnetization switching\\}
	% Force line breaks with \\
	
	\author{M. Kohlmann}
	\affiliation{Institute of Physics, University of Greifswald, Greifswald, Germany}
	\author{L. Vollroth}
	\affiliation{Institute of Physics, University of Greifswald, Greifswald, Germany}
	\author{K. J\"ackel}
	\affiliation{Institute of Physics, University of Greifswald, Greifswald, Germany}
	\author{K. Hovorakova}
	\affiliation{Faculty of Mathematics and Physics,Department of Chemical Physics and Optics, Charles University, Prague, Czech Republic}
	\author{E. Schmoranzerova}
	\affiliation{Faculty of Mathematics and Physics,Department of Chemical Physics and Optics, Charles University, Prague, Czech Republic}
	\author{K. Carva}
	\affiliation{Faculty of Mathematics and Physics, Department of Condensed Matter Physics, Charles University, Prague, Czech Republic}
	\author{D. Hinzke}
	\affiliation{Department of Physics, University of Konstanz, Konstanz, Germany}
	\author{U. Nowak}
	\affiliation{Department of Physics, University of Konstanz, Konstanz, Germany}
	\author{M. M{\"u}nzenberg}
	\affiliation{Institute of Physics, University of Greifswald, Greifswald, Germany}
	\author{J. Walowski}
	\email{jakob.walowski@uni-greifswald.de}
	\affiliation{Institute of Physics, University of Greifswald, Greifswald, Germany}
	
		\date{\today}% It is always \today, today,
	
\begin{abstract}
All-optical helicity-dependent magnetization switching (AO-HDS) is the quickest data recording technique using only ultrashort laser pulses. FePt grains provide an ideal platform for examining the interaction of effects conducting magnetization switching. We identify the magnetic circular dichroism (MCD) and the inverse Faraday effect (IFE) as the primary switching forces. Ultrafast photon absorption rapidly elevates electron temperatures, quenching magnetization. The MCD's helicity-dependent absorption ensures distinct electron temperatures, holding a finite switching probability by generating different spin noise rates in each spin channel. The IFE induces a magnetic moment, enhancing this probability. We present ultrashort laser pulse (<200 fs) AO-HDS experiments in the near-infrared spectral range from 800 nm to 1500 nm, demonstrating a correlation between switching efficiency and absorbed energy density. Elevating electron temperatures to the Curie point enables the IFE to induce a magnetic moment for deterministic switching in the quenched magnetization state. Unlike in films or multilayers, where domain wall motion and domain growth govern the switching process, increasing the MCD in nanometer-sized grains does not enhance switching efficiency. Electrons around Curie temperature typically reach increased switching rates for higher induced magnetization generated by the IFE. The MCD sets the necessary switching condition, separating electron temperatures. The IFE generates a magnetic moment, directing spins toward the desired orientation and improving switching efficiency. Every laser pulse initiates a new switching probability for each grain, increasing the role of direction indication by the IFE. Stronger absorption assures higher induced magnetization at low switching fluences. 
\end{abstract}
	
	
\maketitle
	
\section{Introduction}
Ultrafast magnetization dynamics remains a topic of ongoing research, recently pushing into the attosecond range \cite{Siegrist.2019, Neufeld.2023}, while, at the femtosecond timescale, the interest has shifted to the search for applications. Especially the deterministic magnetization control enables the exploration of all-optical magnetization switching, which is a promising application for memory storage and magnetic sensors. Several switching mechanisms in metals have been recently investigated in numerous publications, splitting the field into all-optical helicity dependent switching (AO-HDS), using circularly polarized laser pulses, and all-optical helicity independent switching (AO-HIS) \cite{Radu.2011, LeGuyader.2012, Lalieu.2017, Verges.2023}, using linearly polarized laser pulses. AO-HDS was first demonstrated on ferrimagnets\cite{Stanciu.2007} and since then has been applied to ferromagnetic materials \cite{Mangin.2014,Lambert.2014,John.2017,Granitzka.2017, Yamamoto.2019,Cheng.2020,Takahashi.2023}. Both techniques open possibilities for more specific switching processes, such as toggle switching \cite{Ignatyeva.2019, Banerjee.2020} and switching by moving domain walls in thin films or layered systems \cite{Liao.2019, Takahashi.2023, Liu.2023, Yoshikawa.2023}. One prominent research aspect is the reduction of laser pulses required to drive the switching process\cite{Yamada.2022}. Additionally, investigations on more complicated magnetic structures like dielectric crystals with a four-fold magnetic anisotropy reveal the possibility of controlling magnetization on ultra-fast time scales \cite{Stupakiewicz.2017, Stupakiewicz.2019}. More recently, Stupakiewicz et al. have started exploring ways to control the magnetic order through the phononic system\cite{Stupakiewicz.2021, Frej.2023}. Recent theoretical calculations even show the potential for switching magnetization in anti-ferromagnetic materials like CrPt \cite{Dannegger.2021}. The state of this development has been published in the review by Wang and Liu in reference\cite{Wang2020} in 2020.
All of these experiments provide insight into the essential factors for ultra-fast magnetization control in specific material systems and their impact on the magnetic ordering process. While in metals, this process requires heating the electron system to the vicinity of the Curie temperature, the switching process in dielectrics typically takes place non-thermally.
	
AO-HDS is a promising approach for magnetic data recording in future generations of magnetic storage devices because of its integration capability into heat-assisted magnetic recording (HAMR) technology in subsequent development steps. In contrast to HAMR, AO-HDS works purely optically, utilizing circularly polarized laser pulses. In the latest picture, for ferromagnetic granular media, the MCD and the IFE are understood as the effects contributing to the magnetization switching in AO-HDS \cite{John.2017}. Research on granular alloys like CoPt or FePt \cite{Weller.2013,Takahashi.2016,Suzuki.2023, Tsai.2023, Tham.2023} confirms that these materials are the perfect choice for memory applications due to their large coercive fields and saturation magnetization, assuring magnetic stability at room temperature. The grains small dimensions down to diameters $\sim 4\,\si{\nano\meter}$ 
and magnetic separation grants single domain switching by switching the whole grains magnetization, enabling bit area densities (AD) around $4\,\si{\tera b \per\square {\centi\meter}}$ \cite{Weller.2014, Suzuki.2023b}.
	
The employment of AO-HDS in magnetic data storage devices for ultrafast recording requires further elucidation of the involved processes. So far, the switching experiments usually have been conducted with laser pulses at wavelengths around $\lambda=800\,\mathrm{nm}$ corresponding to photon energies $E=1.55\,\si{e\volt}$. Steil et al. investigate wavelength dependence towards higher photon energies, generally revealing increasing threshold fluences for switching in this regime\cite{Steil.2011}. While Hennecke et al. even employ high energy photons in the ultraviolet spectral range for their investigations\cite{Hennecke.2024}. Stiehl et al. investigate switching enhancement in chromium and manganese doped FePt\cite{Stiehl.2024}. The calculations published in \cite{John.2017,Berritta.2016} point to enhanced switching efficiency for smaller photon energies due to the increasing difference in both the complex part of the refractive index and the IFE coefficient for both helicities.
	
In this publication, we study the wavelength-dependent switching efficiency of granular FePt media in AO-HDS experiments. We present data obtained by switching magnetization using laser pulses with central wavelengths  $\lambda_{\mathrm{c}}=800\,\si{\nano\meter}$ and in the range of $\lambda_{\mathrm{c}}=1200\,\si{\nano\meter} - 1500\,\si{\nano\meter}$, gaining a deeper insight into the switching efficiency dependence on photon energy. Our data clearly shows a decreasing switching efficiency for smaller photon energies. We explain this outcome by less effective heating of the electron system towards the vicinity of the Curie Temperature due to a decreasing overall absorption for those wavelengths.

\section*{Experiments}
\begin{figure}%[H]
	\centering
	\includegraphics[width=0.9\linewidth]{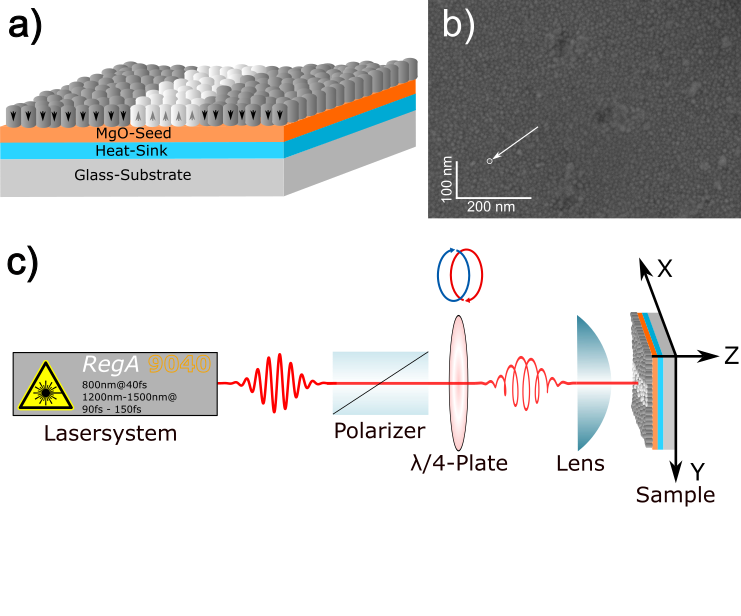}
	\caption{a) Schematic of the sample structure used in the experiments. The magnetic grains, shown in white and grey, populate the MgO seed-layer unordered. The arrows visualize the grains strong out-of-plane magnetization anisotropy.
		b) SEM picture of the sample surface with an average grains size $\sim10\,\si{\nano\meter}$.
		c) Experimental setup schematic. linearly polarized Laser pulses from a Ti:Sapphire laser are circularly polarized by a quarter wave plate and subsequently focused by a lens to a spot size of $40\,\si{\micro\meter}$, illuminating a sample mounted to a transitional stage.}
	\label{fig:experiment_schematic}
\end{figure}

We conduct all switching experiments on FePt $L1_{0}$ granular media previously examined in\cite{John.2017}. The individual grains with diameters around $10\,\si{\nano\meter}$ are embedded into a non-magnetic protective carbon matrix, which separates the grains and their magnetization. They are grown by sputter deposition on an MgO seed layer with a NiTa heat sink on glass substrates, see schematic and electron microscope image in Fig. \ref{fig:experiment_schematic}a) and b) respectively. The grains have a strong perpendicular magnetic anisotropy with coercive fields of $\mu_0 H_c = 4\,\si{\tesla}$ and saturation fields of $\mu_0 H_s = 6\,\si{\tesla}$ \cite{Weller.2013}. We prepare the samples for switching experiments by applying heating above the Curie temperature $T_C=730\,\si{\kelvin}$ and cooling to room temperature cycles in an ultra-high vacuum without an applied magnetic field. This procedure ensures an initial state with a $50\% - 50\%$ up-down magnetic configuration of the grains.
%Cooling down with a 400 \si{\milli\tesla} applied field -> magnetized sample

Fig. \ref{fig:experiment_schematic}c) displays the experimental setup schematics. A Ti:Sapphire Regenerative Amplifier RegA 9040 seeded by a Mira Ti:Sapphire mode-locked oscillator (Coherent) provides the laser pulses with a repetition rate at $250\,\si{\kilo\hertz}$, central wavelength $\lambda_{\mathrm{c}} = 800\,\si{\nano\meter}$, full-width half-maximum $\mathrm{FWHM}\approx 35\,\si{\nano\meter}$ spectral bandwidth and pulse duration $\tau_P = 45\,\si{\femto\second}$ after compression. Time intervals of $4\,\si{\micro\second}$ between pulses ensure that magnetization dynamics processes occurring on the nanosecond time scale are complete before the next pulse arrives. The pulses are tuned by an optical parametric amplifier, Coherent OPA 9850, to the wavelength range $\lambda_{\mathrm{c}} = 1200\,\si{\nano\meter} - 1500\,\si{\nano\meter}$ with a longer pulse duration $\tau_P = 90\,\si{\femto\second} - 150\,\si{\femto\second}$ estimated from the time bandwidth product. A lens focuses the beam spot to a diameter $d=40\,\si{\micro\meter}\pm 1\,\si{\micro\meter}$, corresponding to a spot size $A_B=1260 \pm 260\,\si{\square{\micro\meter}}$, at the sample surface, determined before every switching event for each wavelength. The power deposited on the sample surface $P = 0 - 60\,\si{\milli\watt}$ corresponds to fluences from $F = 3\,\si{\milli\joule\per\square{\centi\meter}} - 20\,\si{\milli\joule\per\square{\centi\meter}}$ and peak intensities from $I_{\mathrm{Peak}} = 5\,\si{\giga\watt\per\square{\centi\meter}} - \sim400\,\si{\giga\watt\per\square{\centi\meter}}$. We set the helicity by a polarizer quarter-wave-plate sequence, using a zero-order wave-plate for $\lambda_{\mathrm{c}} = 800\,\si{\nano\meter}$ and an achromatic wave-plate AQWP05M-1600 (Thorlabs Inc) for the $\lambda_{\mathrm{c}} = 1200\,\si{\nano\meter} - 1500\,\si{\nano\meter}$ range. The determined helicity of the circular laser pulses is greater than $80\%\pm 10\%$.

The motorized $xyz$-transition sample holder stage enables transitions in $xy$-direction to move the laser beam across the sample surface and in $z$-direction for spot size fine adjustment in $200\,\si{\nano\meter}$ steps and speeds up to $20\,\si{\milli\meter\per\second}$. Combining the stage parameters with the pulse repetition rate, we deposit $\sim 500$ pulses per $1/e^2$ beam diameter at every spot. This number of pulses ensures reaching the possible magnetization saturation for each helicity \cite{John.2017}.

We image the switched magnetic areas using a magneto-optical Kerr microscope with polar sensitivity, recording the images by a Zyla 5.5 camera (Andor - Oxford Instruments). A Zeiss ec epiplan x 50 with a numeric aperture $NA = 0.75$ objective with a $\lambda = 525\,\si{\nano\meter}$ light source SOLIS-525C (Thorlabs Inc) configured for K\"ohler-illumination define the amplification. An additional quarter-wave plate employed in the crossed polarizer configuration compensates for the Kerr ellipticity in the light reflected from the sample surface. All evaluated images are averages over 100 captured frames, subtracting the background using a clean demagnetized area to eliminate illumination inhomogeneities, obtaining maximum magnetic contrast. 

\begin{figure}%[H]
	\centering
	\includegraphics[width=0.45\linewidth]{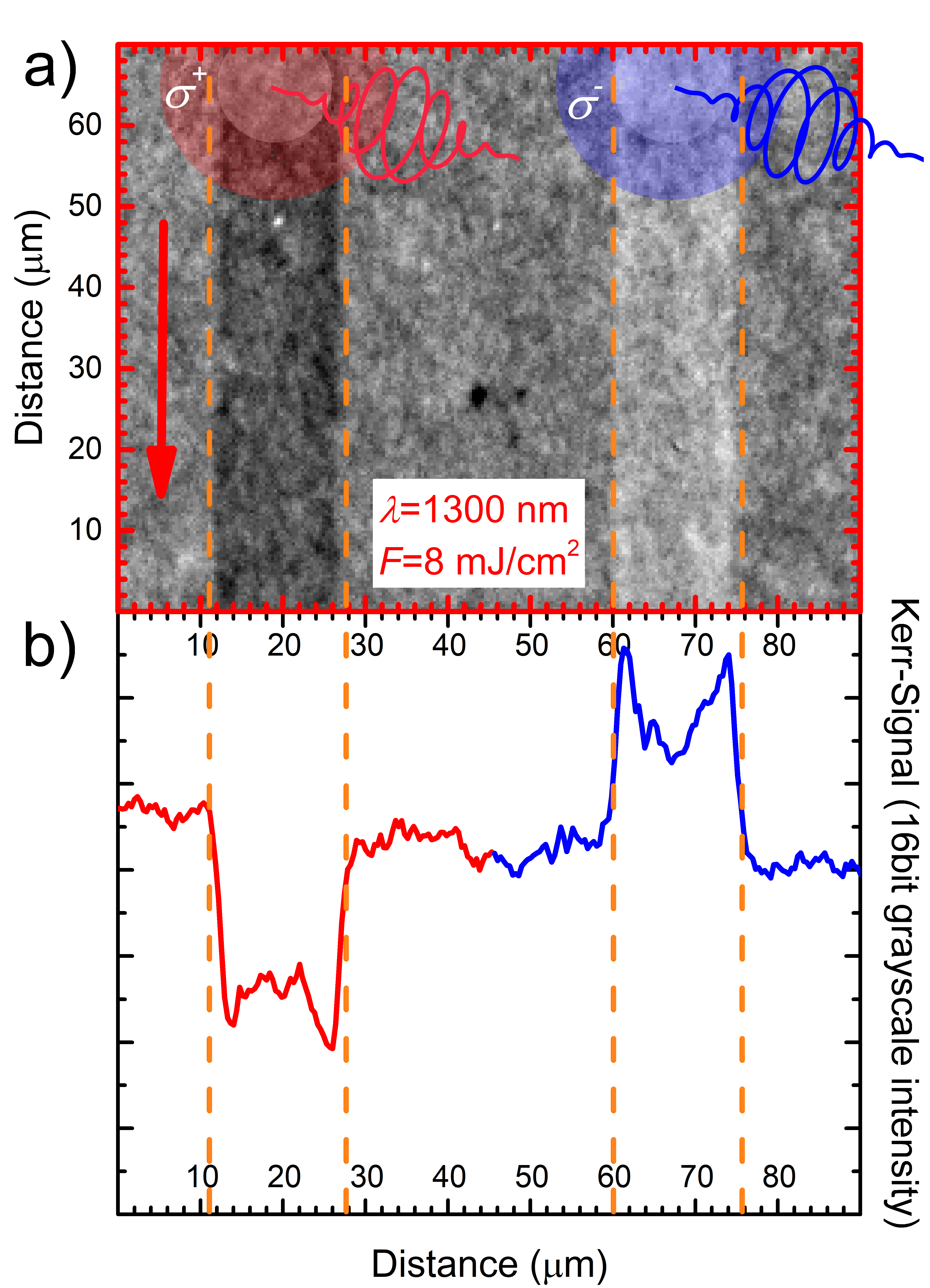}
	\caption{a) Kerr-microscopy image (16-bit gray scale values) of a demagnetized sample with lines switched by $\lambda_{\mathrm{c}} = 1300\, \si{\nano\meter}$ pulses at fluence $F = 8\,\si{\milli\joule\per\square\centi\meter}$. The red arrow indicates the sample transition direction for the extracted profiles b) by integrating the area marked by the red rectangular box for signal-to-noise ratio reduction.}
	\label{fig:sample_line}
\end{figure}

After imaging, we extract the profiles across the written lines from the Kerr images to obtain the Kerr signal and switched area size and evaluate the switching efficiency. Fig. \ref{fig:sample_line} outlines the procedure for lines written employing the wavelength $\lambda_{\mathrm{c}}=1300\,\si{\nano\meter}$ at the fluence $F\approx 8\,\si{\milli\joule\per\square{\centi\meter}}$. Both lines in the Kerr-microscope magnetic greyscale image show magnetization switched along the direction of the red arrow using opposite helicities starting from a demagnetized state. The spots indicate the laser beam, red ($\sigma^{+}$) and blue ($\sigma^{-}$). We obtain the intensity value profiles from the image, averaging the magnetization contrast within the area edged in red over multiple pixel rows to improve the signal-to-noise ratio (SNR) by defining the switched area, where the Kerr signal exceeds the local noise level. Two extracted representative profiles showing the cross-section Kerr signal are plotted in Fig. \ref{fig:sample_line}b). We define the areas switched using right-hand helicity, $\sigma^{+}$ as the darker areas, decreasing the Kerr signal, and the areas switched using left-hand helicity, $\sigma^{-}$ as the brighter areas, increasing the Kerr signal. The resulting Kerr signal is proportional to the switched magnetization or the switching rate. At this point, first, we observe a nonuniform magnetization across the lines, indicating a lower switching rate in the center than on the edges. Second, the profile width is $\sim20\,\si{\micro\meter}$, only one-half of the beam spot diameter. We will refer to these occurrences throughout the discussion.

\section*{Discussion}

\begin{figure}%[H]
	\centering
	\includegraphics[width=0.8\linewidth]{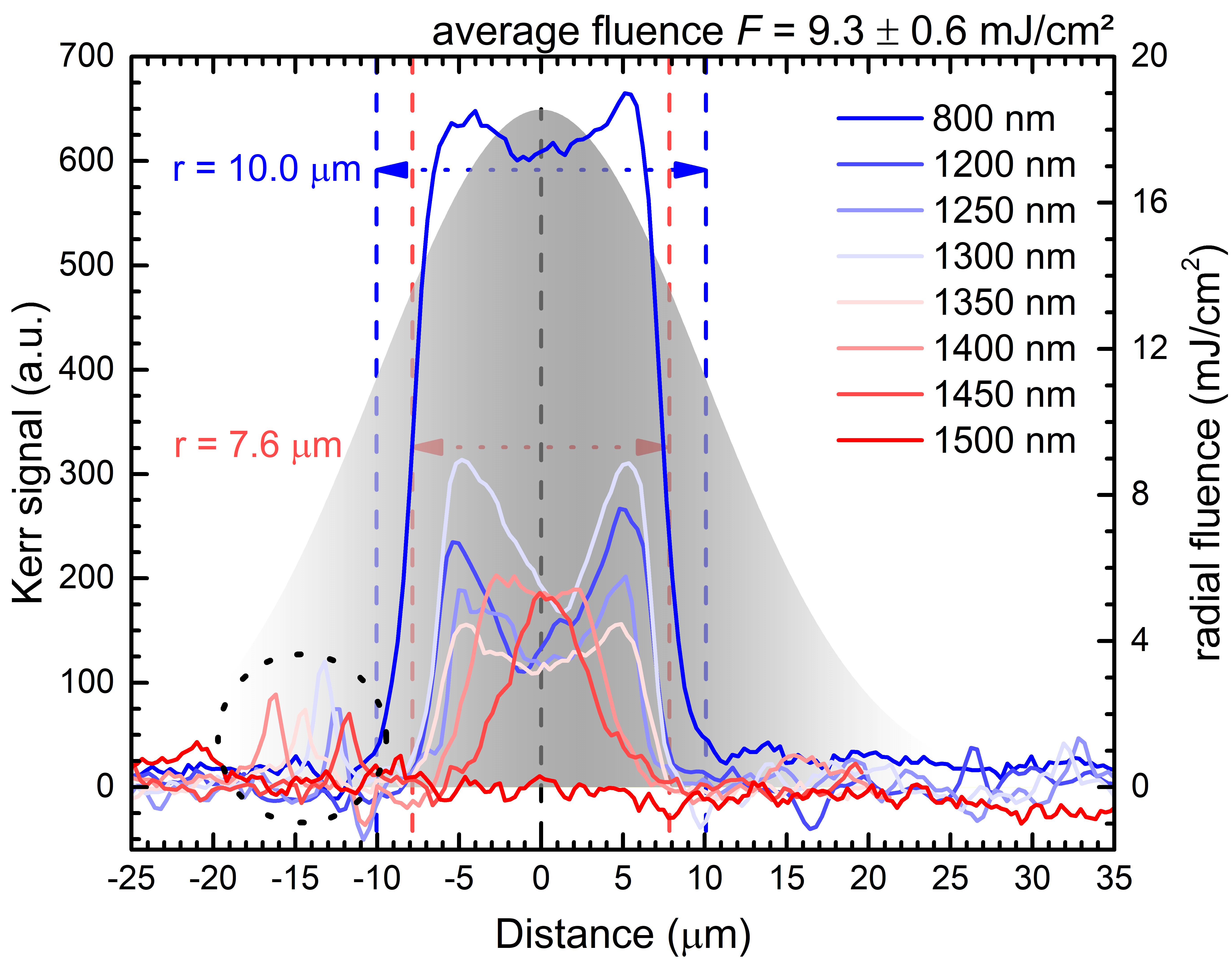}
	\caption{Extracted line profiles switched using $\sigma^{-}$ helicity for all wavelengths $\lambda_{\mathrm{c}} = 800\,\si{\nano\meter} - 1500\,\si{\nano\meter}$ at a deposited average fluence $F=9.3\pm0.6\,\si{\milli\joule\per\square{\centi\meter}}$ scaled on the left y-axis. The grey-shaded Gaussian function in the background represents the radial fluence distribution scaled on the right y-axis. The black dashed line marks the center of the written lines. The blue and red dashed lines define the line widths for switching with $\lambda_{\mathrm{c}} = 800\,\si{\nano\meter}$ and $\lambda_{\mathrm{c}} = 1450\,\si{\nano\meter}$ pulses respectively. The black dotted circle points out peaks originating from surface distortions.} 
	\label{fig:profiles}
\end{figure}

In Fig. \ref{fig:profiles}, we examine the switching characteristics for the investigated wavelengths using an average irradiation fluence $F=9.3\pm0.6\,\si{\milli\joule\per\square\centi\meter}$. The grey-shaded function in the background approximates the radial fluence distribution for Gaussian-shaped laser spots with a radius $r = 20\,\si{\micro\meter}$ scaled on the right y-axis. The profiles obtained after each switching process using $\sigma^-$ helicity for all applied wavelengths are shifted to 0, marking the initial demagnetized state, while the positive values represent the switched magnetization for direct comparison. This routine corrects for the minor brightness deviations ($<5\%$) and illumination offsets between the Kerr images. Without loss of generality, we only discuss the $\sigma^-$ data because the profiles for both helicities are symmetric for an initial demagnetized sample state. We added the profiles showing this symmetry and the background noise in the Supplementary materials section in Figs. \ref{fig:symswitch1} to \ref{fig:symswitchaverage}. The profile obtained using $\lambda_{\mathrm{c}}=800\,\si{\nano\meter}$ pulses stands out, the Kerr signal is twice as high over the whole profile, and the switched magnetization area is slightly larger, corresponding to a radius $r\approx 10.0\,\si{\micro\meter}$, indicated by the blue dashed lines. While the switched areas in all other profiles correspond to radii $r\lesssim 7.6\,\si{\micro\meter}$, marked by the red dashed lines, are only around $1/3$ of the laser spot radius. Both quantities, the Kerr signal and the switched area, decrease by shifting the wavelength towards $\lambda_{\mathrm{c}} = 1500\,\si{\nano\meter}$. This behavior hints at an elevated switching efficiency for $\lambda_{\mathrm{c}}=800\,\si{\nano\meter}$ pulses compared to the other employed wavelengths.

We find three switching stages at this fluence within the investigated spectrum. The $\lambda_{\mathrm{c}} = 1500\,\si{\nano\meter}$ pulses do not initiate any switching process yet. Both shorter wavelengths, $\lambda_{\mathrm{c}}=1450\,\si{nm}$ and $\lambda_{\mathrm{c}}=1400\,\si{nm}$ achieve their maximum switching capability indicated by the peak and the plateau in the profile, respectively. While all profiles, for the wavelengths $\lambda_{\mathrm{c}}\leq 1300\,\si{\nano\meter}$ show a drop in the center of the switched area marked by the dashed black line in Fig. \ref{fig:profiles} and thus, a reduced switching rate. 
The higher energy concentration in the beam center of a Gaussian beam profile generates nonuniform electron temperatures and varying magnetization dynamics across the switched area. While electrons at the edges reach temperatures above $T_{\mathrm{C}}$ for a brief period $\leq 100\,\si{\femto\second}$, subsequently returning into the ferromagnetic state after the laser pulse ends, the electrons in the center reach temperatures twice as high as $T_{\mathrm{C}}$, remaining in this elevated state for up to around $\approx 1\,\si{\pico\second}$.
First, those higher temperatures cease the magnetic order and cause a more randomized magnetic state. Second, electrons at the edges have a reduced magnetization relaxation time and reverse their magnetization towards the 'up' state, generating demagnetizing fields that reduce the switching probability for still-hot electrons in the beam center, forcing a lower switching rate. The outcome is consistent with magnetization dynamics and the extracted electron temperatures presented in \cite{Mendil.2014}. Within our dataset, this switching rate reduction differs for the employed wavelengths but does not show any distinguished trend for the wavelength range because the switching procedure is more complex, containing additional wavelength-dependent effects. Due to lower overall electron temperatures, the longer wavelengths do not achieve switching rate suppression in the profile center. They merely achieve electron temperatures sufficient for magnetization switching in the center. Lastly, the slight profile asymmetries originate from laser beam astigmatism, while the peaks encircled by the black dots, showing in some of the profiles, originate from impurities on the sample surface. In the following, we extract the maximum Kerr signal and the switched areas from all profiles for a systematic analysis to discuss the relation between the amount of deposited energy required for magnetization switching and to evaluate the resulting efficiency. For this comparison, we compare the Kerr signals to those of fully magnetized samples $M_{\mathrm{S}}$ and calibrate the switched magnetization ratio $\Delta M/M_{\mathrm{S}}$ using the same procedure as in \cite{John.2017}.

\begin{figure}%[H]
	\centering
	\includegraphics[width=1.\linewidth]{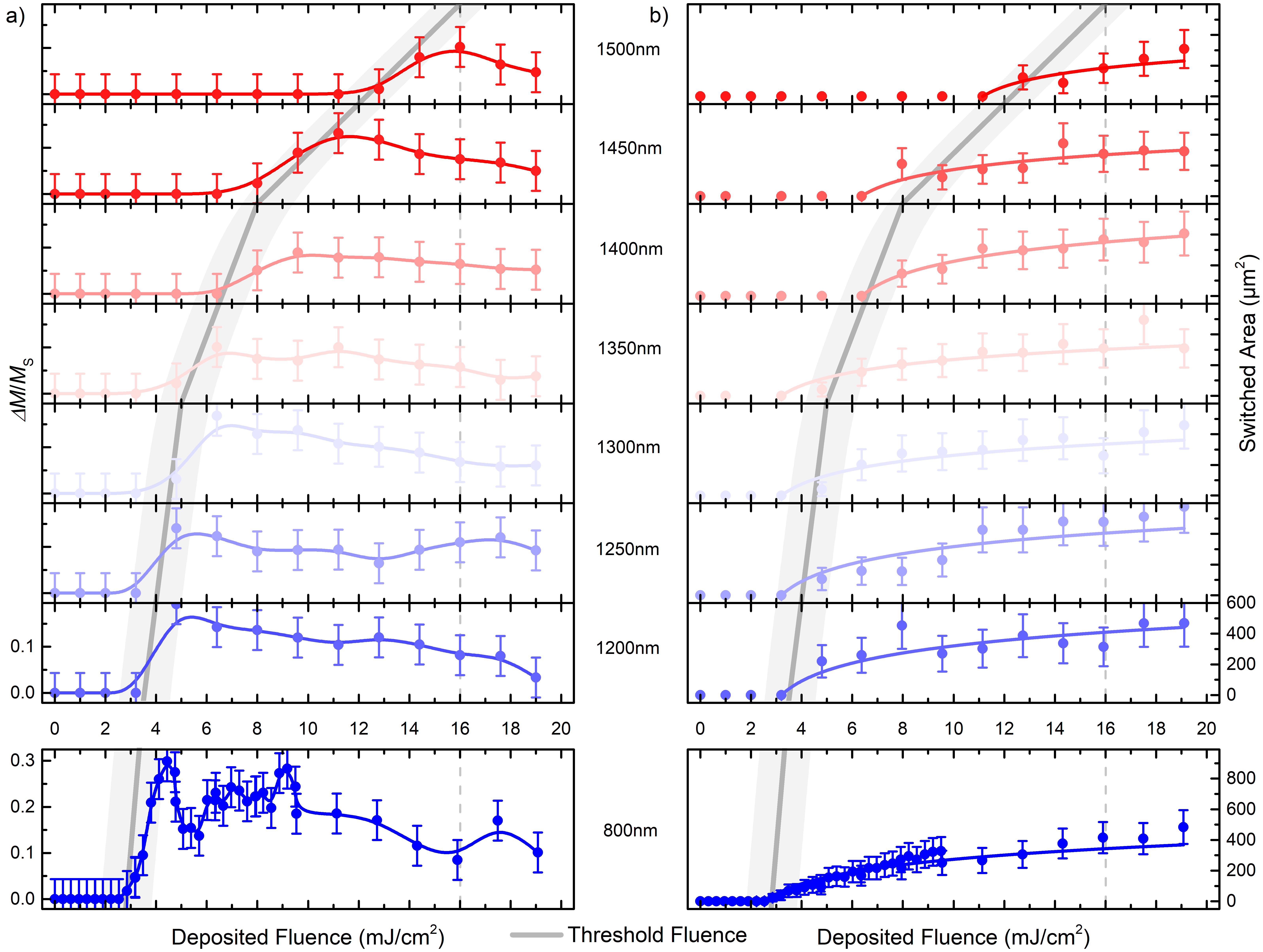}
	\caption{a) Switched magnetization $\Delta M/M_{\mathrm{S}}$ (lines are a guide to the eye) and b) switched area $A_{\mathrm{S}}$ (lines are linear fits) versus deposited laser fluence for the investigated spectrum. a) $\Delta M/M_{\mathrm{S}}$ reaches a peak at optimum fluence, then the switching efficiency decreases for increasing fluences. b) $A_{\mathrm{S}}$ increases with the deposited laser fluence. The gray lines (light gray exterior outlines the error) trace the spectral threshold fluence dependence increasing with the wavelength.}
	\label{fig:wavelength_dependence}
\end{figure}

Fig. \ref{fig:wavelength_dependence} displays the maximum switched magnetization $\Delta M/M_{\mathrm{S}}$ a) and the switched area b), plotted as functions of deposited irradiation fluence. The data are arranged from $\lambda_{\mathrm{c}} = 1500\,\si{\nano\meter}$ in the top panel to $\lambda_{\mathrm{c}} = 800\,\si{\nano\meter}$ in the bottom panel. Note that the scale range for the lower panel is larger, allowing for constant scaling throughout the panels. We identify a threshold laser fluence initiating the switching process within the investigated spectrum, increasing for longer wavelengths, plotted in the background by the grey lines across the panels as a guide to the eye. The light grey areas outline the corresponding threshold fluence error. Each wavelength has an optimum fluence providing a maximum $\Delta M/M_{\mathrm{S}}$, usually around $1\,\si{\milli\joule\per\square{\centi\meter}}$ to $3\,\si{\milli\joule\per\square{\centi\meter}}$ above the threshold fluence. Further, the $\Delta M/M_{\mathrm{S}}$ peak decreases by up to $\sim70\%$ within the spectrum from $\lambda_{\mathrm{c}} = 800\,\si{\nano\meter} - 1500\,\si{\nano\meter}$. Finally, at fluences exceeding the optimum switching point, $\Delta M/M_{\mathrm{S}}$ gradually decreases due to a reduced overall switching efficiency for all applied wavelengths.

This decreasing trend in the switching rate, when fluences exceed the optimum switching fluence, coincides with multiscale switching rate model simulations for antiferromagnets published in reference \cite{Dannegger.2021}. They reveal that both processes, the differing heat entry through the MCD and the magnetization induced by the IFE, contribute to magnetization switching and can trigger a switching rate independently. However, the heat entry alone generates exceptionally low switching probabilities at fluences below the material melting point. Therefore, the IFE, which scales with the laser fluence, significantly increases the switching probability around the critical temperature. Higher temperatures increase the spin system disorder and undermine the deterministic effect of the IFE. Hence, more heat generates higher electron temperatures, increases the spin noise, suppresses the impact of the IFE, and reduces the switching rate. Additionally, the emerging demagnetizing fields diminish the switching rates see Fig. \ref{fig:profiles}. All these occurrences superpose to nontrivial amplifying and damping trends, depending on wavelength and fluence.

% in the below section, the calculated value for the switched area is: 1256 \pm 251.38 \mum^2 

The switched area $A_{\mathrm{S}}$, in Fig. \ref{fig:wavelength_dependence}b) explains the function of heat entering the magnetic material for switching. After reaching the threshold fluence, $A_{\mathrm{S}}\sim\ln(F)$ for all wavelengths, outlined by the fitted lines. The maximum $A_{\mathrm{S}}$ declines with increasing wavelength and remaining below $600\,\si{\square{\micro\meter}}$, does not reach half of the laser spot size for the investigated spectrum, which is consistent with the Kerr signal behavior. While $A_{\mathrm{S}}$ extends linearly with the introduced amount of heat as the threshold fluence reaches the beam edges, the Kerr signal reaches a Maximum and decreases slightly for growing fluence. This behavior suggests an optimum amount of introduced heat and the magnetic moment induced by the IFE for each wavelength and a superior switching efficiency for the $\lambda_{\mathrm{c}}=800\,\si{\nano\meter}$ laser pulses compared to the rest of the investigated spectrum. Our further analysis focuses on the effects acting during the absorption of circularly polarized laser pulses and their balancing to trigger magnetization reversal.

\begin{figure}%[H]
	\centering
	\includegraphics[width=1.\linewidth]{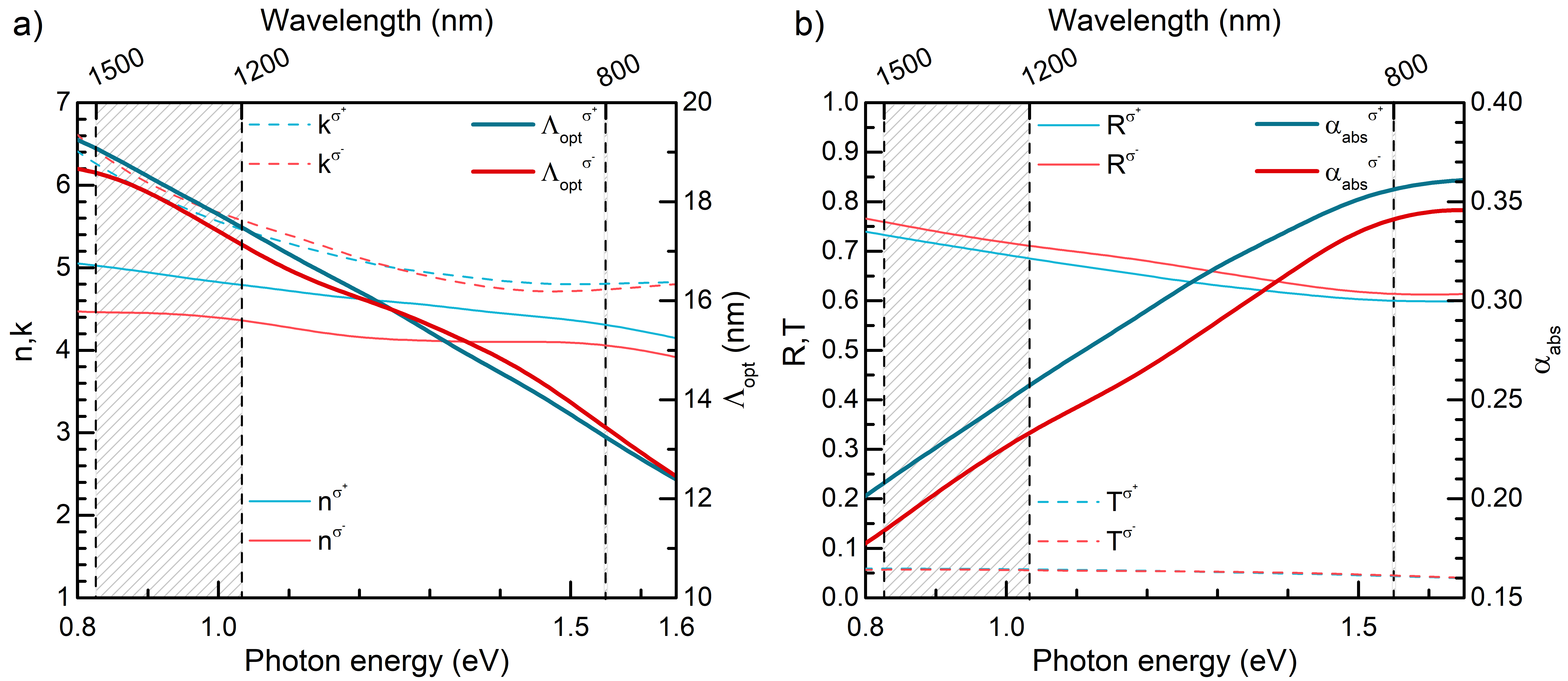}
	\caption{a) Ab-initio calculated complex refractive index $n+ik$ with the real and imaginary part, respectively, calculated for both helicities and the resulting penetration depth $\Lambda_{\mathrm{opt}}$. b) Reflectivity $R$ and transmission $T$, and the resulting absorption $\alpha_{\mathrm{abs}}$ for a $10\,\mathrm{nm}$ thick FePt layer.}
	\label{fig:optical_parameters}
\end{figure}

Accordingly, we examine the optical properties derived from the helicity-dependent refractive index for FePt with $\mathrm{L1_0}$ crystalline structure at normal incidence $(n^{\sigma^{\pm}}+ik^{\sigma^{\pm}})^2 =\varepsilon_{xx}\pm i\varepsilon_{xy}$\cite{Oppeneer.2001}, by calculating the dielectric tensors $\varepsilon_{ij}$ using the Kubo linear response theory. 
For this, we perform density functional theory (DFT) calculations based on the full-potential linear augmented plane wave (FP-LAPW) method, as implemented in the band structure program ELK\cite{ELK.2024}, with local spin-density approximation (LSDA) determining the exchange-correlation potential, and including spin-orbit coupling (SOC). Sampling the full Brillouin zone by $10^4$ wave vector points is usually sufficient for magnetic anisotropy energy calculations in FePt \cite{Burkert.2005}. We confirm this by verifying the convergence of the wave vector mesh density.

Both refractive index components, real $n$ and imaginary $k$ in Fig. \ref{fig:optical_parameters} a) slightly increase with the wavelength, $n$ showing a more distinct helicity disparity (left scale). The resulting penetration depth $\Lambda_{\mathrm{opt}}=\frac{\lambda}{4\pi\cdot k}$, ranges between $13\,\si{\nano\meter}$ for $\lambda_{\mathrm{c}}=800\,\si{\nano\meter}$ and $\sim 19\,\si{\nano\meter}$ for the longer wavelengths (right scale). The overall penetration depth exceeds the FePt grain size, while the difference due to helicity remains below $0.5\,\si{\nano\meter}$ for the whole spectrum. The transfer matrix method provided within the package \cite{Byrnes2021} calculates the reflectivity $R$ and the transmission $T$ through the FePt layer for a layer stack FePt/MgO/NiTa at normal incidence. This data delivers the corresponding absorption $\alpha_{\mathrm{abs}}$ plotted in Fig. \ref{fig:optical_parameters} b), (right scale). The overall absorption $\alpha_{\mathrm{abs}}$ drops by a factor $\geq 1.5$ from $\alpha_{\mathrm{abs}}(\lambda_{\mathrm{c}}=800\,\si{\nano\meter})=0.35$ to $\alpha_{\mathrm{abs}}(\lambda_{\mathrm{c}}\geq 1200\,\si{\nano\meter})<0.25$. Simultaneously, the helicity-dependent absorption difference increases for the longer wavelengths. This results in an MCD increasing from $\sim 2\%$ for $\lambda_{\mathrm{c}}=800\,\si{\nano\meter}$ to between $5\%$ and $6\%$ for $\lambda_{\mathrm{c}}\geq 1200\,\si{\nano\meter}$.

These calculations agree with the tendency observed in the experimental data. The enhanced absorption of shorter wavelengths conducts higher electron temperatures more efficiently, yielding a lower threshold fluence for switching. Electron temperatures close to $T_{\mathrm{C}}$ are crucial. At this point, the temperature separation by the MCD inducing a switching probability for one spin channel works most efficiently. In addition, at those temperatures, a quenched magnetization \cite{Mendil.2014} facilitates switching by the low induced magnetization of the inverse Faraday effect $\Delta M^{\mathrm{ind}}$. Heating the electron system far above $T_{\mathrm{C}}$ is disadvantageous because the growing thermal disorder diminishes the impact of $\Delta M^{\mathrm{ind}}$ induced by the IFE. Further, the elevated switching rates achieved by $\lambda_{\mathrm{c}}=800\,\si{\nano\meter}$ laser pulses imply a significant contribution of the IFE to the switching process in addition to the MCD. The ratio $\Delta M/M_{\mathrm{S}}$ requires a review as a function of absorbed fluence. Accordingly, we multiply the deposited fluence in Fig. \ref{fig:wavelength_dependence} a)  by the absorption from Fig. \ref{fig:optical_parameters} b), and discuss it in the context of $\Delta M^{\mathrm{ind}}$ calculated from the available IFE data. 

\begin{figure}%[H]
	\centering
	\includegraphics[width=1.\linewidth]{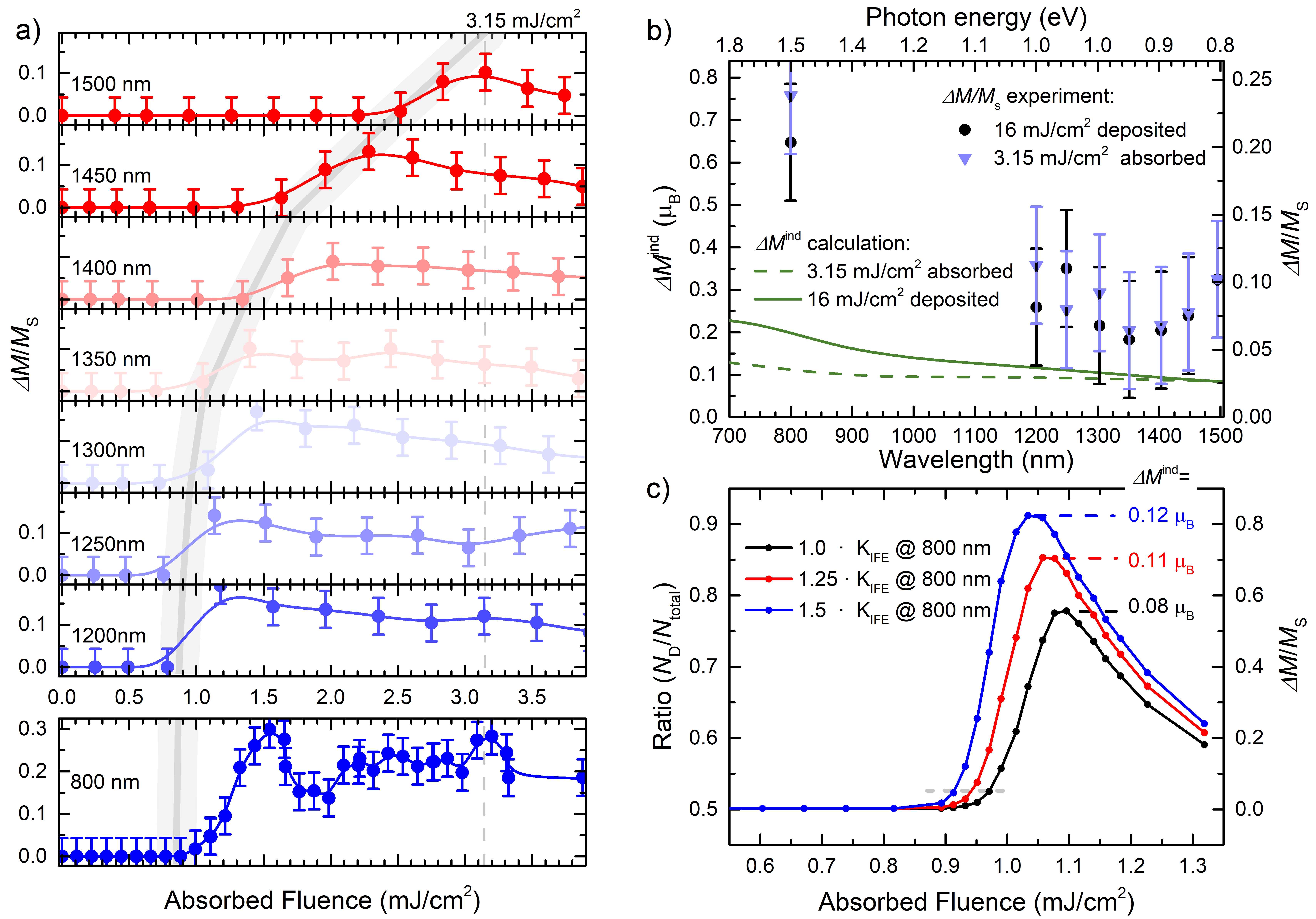}
	\caption{a) Kerr signal vs. absorbed laser fluence. The grey line marks the threshold fluence. b) Magnetization $\Delta M^{\mathrm{ind}}$ induced by the IFE (left scale) for both helicities (blue and red lines) for the deposited fluence of $16\,\si{\milli\joule\per\square{\centi\meter}}$ (solid) and for the absorbed fluence of $1.65\,\si{\milli\joule\per\square{\centi\meter}}$ (dashed). The extracted Kerr signal (right scale) for the fluence $16\,\si{\milli\joule\per\square{\centi\meter}}$ deposited (black dots) and for $1.65\,\si{\milli\joule\per\square{\centi\meter}}$ absorbed (blue stars). c) Multi-scale modeling of the switching ratio for $50\%$ of grains initially in the magnetic "down" state $N_{\mathrm{D}}/N_{\mathrm{total}}=0.5$ after illumination at the absorbed fluence with 30 pulses for three different IFE quantities.}
	\label{fig:absorbed_merged}
\end{figure}

The revised data in Fig. \ref{fig:absorbed_merged} a) confirm a steadily increasing spectral threshold fluence level. It stays around $1\,\si{\milli\joule\per\square{\centi\meter}}$ across a wide wavelength range up to $\lambda_{\mathrm{c}}=1400\,\si{nm}$. For $\lambda_{\mathrm{c}}\geq 1450\,\si{nm}$, the threshold fluence doubles due to slightly enlarged pulse durations and lower peak intensities.
The grey line across the panels marks this trend. The threshold fluences agree with findings by Steil et al. \cite{Steil.2011} for GdFeCo for wavelengths $\lambda_{\mathrm{c}}=400\,\si{nm} - 800\,\si{nm}$. 
$\Delta M/M_{\mathrm{S}}$ reaches a maximum at $\sim0.5\,\si{\milli\joule\per\square{\centi\meter}}$ above the threshold fluence and decreases gradually for increasing fluences within the spectrum. Gorchon et al. \cite{Gorchon.2016} propose an all-optical helicity-dependent switching model based exclusively on MCD. They conclude that an effect of merely $0.5\%$ is sufficient to switch magnetization. The switching condition is obtaining electron temperatures close to $T_{\mathrm{C}}$. Then, electron temperatures below $T_{\mathrm{C}}$ for the spins with the lower absorption do not initiate any switching process, while exceeding $T_{\mathrm{C}}$ for the spins with higher absorption leads to thermal fluctuation. Within this model, a larger MCD increases the switching probability. But the temperature window for switching remains small at $5\%$ around $T_{\mathrm{C}}$. Simultaneously, Ellis et al. \cite{Ellis2016} calculated switching probabilities based on a reptation-like model. They identify the MCD as the main force driving the switching process because the IFE generates magnetic fields that are too small and too short to have an impact. Experiments where applied fields around $100\,\si{\milli\tesla}$ suppress switching published in reference \cite{Lambert.2014} support their conclusion. We also observe the maximum of $\Delta M/M_{\mathrm{S}}$ around $T_{\mathrm{C}}$. The MCD grows from $2\%$ for $\lambda_{\mathrm{c}}=800\,\si{\nano\meter}$ to above $6\%$ for $\lambda_{\mathrm{c}}=1500\,\si{\nano\meter}$, but the data does not show an increased switching probability. Further, raising electron Temperature by more than $5\%$ above $T_{\mathrm{C}}$ by doubling the fluence does not completely demagnetize the illuminated area. The switched magnetization decreases gradually with growing fluence, and material melting sets in before switching effects deplete. Therefore, we estimate the extent to which the IFE amplifies the process once the MCD sets the necessary condition for switching and how both effects interact. We discuss the contribution of the IFE to the switching process by calculating the induced magnetization $\Delta M^{\mathrm{ind}}$ and compare it with the experimental data:
\begin{equation}
	\Delta M^{\mathrm{ind}} = K_{\mathrm{IFE}} \cdot \frac{I}{c}.
	\label{eq:magnetization}
\end{equation}
We use $K_{\mathrm{IFE}}$ values in units of $\si{\per\tesla}$ from reference \cite{Berritta.2016, John.2017}, derived using the quantum-theoretical description of magnetic polarization \cite{Battiato.2014}. $I$ denotes the absorbed laser pulse intensity from experiments, and $c$ is the speed of light. $\Delta M^{\mathrm{ind}}$ has the dimension $\si{\micro_{B}}$ per unit cell. The magnetic moment per unit cell of FePt is $3.24\,\si{\micro_B}$.

At threshold fluence, electron temperatures elevate close to $T_{\mathrm{C}}$ and quench $M_{\mathrm{S}}$ to $\sim 20\%$ of its room temperature value \cite{Mendil.2014}. Simultaneously, the induced magnetization $\Delta M^{\mathrm{ind}}\approx 0.05\,\si{\micro_B}$ stays below $2\%$ of the FePt magnetic moment, leaving the temperature difference between opposite spin states generated by the MCD as the only mechanism, generating a switching probability. Increasing the fluence from this point towards the optimum switching point, we expect a larger MCD to develop more effective switching, clearly separating the electron temperatures of both spin states, because in this range $\Delta M^{\mathrm{ind}}$ remains small. Wavelengths from $\lambda_{\mathrm{c}}=1200\,\si{\nano\meter}$ to $\lambda_{\mathrm{c}}=1500\,\si{\nano\meter}$, generate the larger MCD, between $5\%$ and $6\%$, but we observe a switching ratio less than $\Delta M/M_{\mathrm{S}}=0.2$. While $\lambda_{\mathrm{c}}=800\,\si{\nano\meter}$ generating the smaller MCD around $2\%$ switches roughly $\Delta M/M_{\mathrm{S}}=0.3$. As the fluence increases,  thermal fluctuations lead to greater demagnetization, but the switching efficiency does not break down immediately. Instead, the switching rate decreases gradually over a wide fluence range. 

In the next step, we add the IFE into consideration. In Fig. \ref{fig:absorbed_merged} b) (left scale), we compare $\Delta M^{\mathrm{ind}}$ for the investigated spectrum, illustrating the switching process enhancement by the IFE for two fluences. First, the deposited fluence $F=16\,\si{\milli\joule\per\square{\centi\meter}}$ (solid line), it corresponds to the absorbed fluence of $F_{\mathrm{abs}}=3.15\,\si{\milli\joule\per\square{\centi\meter}}$, and simultaneously achieves the maximum $\Delta M/M_{\mathrm{S}}$ for $\lambda_{\mathrm{c}}=1500\,\si{\nano\meter}$. For the remaining wavelengths, this fluence corresponds to larger $F_{\mathrm{abs}}$, far beyond the maximum $\Delta M/M_{\mathrm{S}}$. In this configuration, $\Delta M^{\mathrm{ind}}(\lambda_{\mathrm{c}}=800\,\si{\nano\meter})=0.2\,\si{\mu_B}$ is about twice as large as for the remaining wavelengths. $\Delta M^{\mathrm{ind}}(\lambda_{\mathrm{c}}\geq 1200\,\si{\nano\meter})<0.12\,\si{\mu_B}$, despite a larger $K_{\mathrm{IFE}}$ absolute value for those wavelengths. Additionally, the lower overall absorption keeps $\Delta M^{\mathrm{ind}}$ at a low level. Second, we compare $\Delta M^{\mathrm{ind}}$ at an equally absorbed fluence $F_{\mathrm{abs}}=3.15\,\si{\milli\joule\per\square{\centi\meter}}$ (dashed line). In this case, $\Delta M^{\mathrm{ind}}$ remains constant with a slight increasing tendency towards $\Delta M^{\mathrm{ind}}(\lambda_{\mathrm{c}}=800\,\si{\nano\meter}) = 0.11\,\si{\mu_B}$.

The experimental data exhibits a qualitatively matching behavior (see Fig. \ref{fig:absorbed_merged} b), right scale) for $\Delta M/M_{\mathrm{S}}$ at the specified fluences. Especially, $\lambda_{\mathrm{c}}=800\,\si{\nano\meter}$ achieves a higher $\Delta M/M_{\mathrm{S}}$ at the lower fluence (blue triangle) than at the higher fluence generating a larger $\Delta M^{\mathrm{ind}}$ because it is closer to the optimum point. In conclusion, enhanced absorption conducts maximum switching efficiency at lower deposited fluences, whereby further rising fluence leads to a gradually decreasing $\Delta M/M_{\mathrm{S}}$ despite a growing $\Delta M^{\mathrm{ind}}$. This behavior arises from electron temperatures exceeding $T_{\mathrm{C}}$ and the resulting thermal fluctuations lasting for picoseconds. Simultaneously, the MCD cannot compensate by inducing enough temperature difference, leaving electron temperatures in grains with opposite magnetization below $T_{\mathrm{C}}$. A closer look at noise correlations on the microscopic level near $T_{\mathrm{C}}$ \cite{Atxitia.2009, Kazantseva.2008} reveals the nuances of these processes. The ideal switching conditions emerge slightly below $T_{\mathrm{C}}$ when susceptibility diverges into two parts, $\chi_{\perp}$ representing magnetization deflection from the initial state direction, and $\chi_{\parallel}$ representing the magnetization vector quenching. In this state, the electrons are susceptive to momentum transfer from polarized photons, inducing a $\Delta M^{\mathrm{ind}}$ to the quenched magnetization, while thermal fluctuations decay before noise correlations introduce disorder.

In contrast, Pt/Co/Pt multilayer film stacks show enhanced magnetization reversal\cite{Yamada2025}. In those systems, domain-wall motion and domain growth govern the switching process. Within this mechanism, the temperature gradient between two magnetization domains controls the switching process, favoring the expansion of the domain whose magnetization absorbs less energy for a given helicity, preserving its magnetization state and making the other domain susceptible to change by thermal fluctuations, aligning with the colder regions. The publication also shows that even a low MCD below $3\%$ is sufficient to obtain a $100\%$ switching efficiency after multiple excitations. Therefore, when magnetic domains are absent, the first laser pulse initiates domain formation. The domains expand according to the set helicity\cite{Parlak2018}. The switching procedure has two steps, where the first $\sim 5$ pulses demagnetize the system forming multi-domains, and the remaining $\lesssim 50$ pulses switch the magnetization in a cumulative process\cite{ElHadri2016}. That's why multilayer helicity-dependent switching experiments reach switching rates up to $90\%$, compared to lower switching rates achieved in single-domain grains. Separated single-domain grains do not experience the support of already magnetically oriented domains. They switch independently, and each heating cycle generates a new switching probability, given by the specific combination of MCD and IFE and the initial magnetization distribution. Under those circumstances, a small $\Delta M^{\mathrm{ind}}$ induced by the IFE enhances the switching process while increasing the MCD within a range below $10\%$ does not significantly increase the probability of switching. 

Lastly, multiscale switching probability simulations employing the procedure developed in \cite{John.2017} mirror the processes driving the electron and spin systems pointed out in the experimental results. The data in Fig. \ref{fig:absorbed_merged} c) show the switching ratio after 30 laser pulses plotted as a function of absorbed fluence. The model reveals a slightly shifted switching threshold fluence compared to the experimental values attributed to the smaller grains used in the simulation. The significant result from this simulation is the increasing impact of the IFE for a constant set of the remaining parameters. A larger $\Delta M^{\mathrm{ind}}$ generates higher switching rates at lower threshold fluences. Although switching rates decrease with rising thermal fluctuations, they do not vanish. The superior $\Delta M/M_{\mathrm{S}}$ outcome in simulations as compared to experiments stems from fluctuations, such as heat dissipation into the heat sink layer or a circular polarization degree below $100\%$ in the experimental procedure.

%magnetic moment for a FePt pair = 3.24\,\si{\mu_B}

%~~~~~~~~~~~~~ conclusion ~~~~~~~~~~~~~

In conclusion, our results demonstrate how the interplay and the combined synergy of the occurring effects influence the efficiency in all-optical helicity-dependent magnetization switching in the near-infrared spectral range. The total energy absorbed by the magnetic grains must generate electron temperatures close to $T_C$ and quench magnetization. In this instance, the MCD is essential. It separates electron temperatures of opposite spins and only elevates one spin species above $T_C$, thus generating a switching probability. However, the switching efficiency does not increase with an elevated MCD. Instead, a greater $\Delta M^{\mathrm{ind}}$ induced by the IFE improves the process. Also, the induced magnetization does not scale directly with $K_{\mathrm{IFE}}$ as expected for the longer wavelengths\cite{Berritta.2016, John.2017}. The contribution of $\Delta M^{\mathrm{ind}}$ unfolds best when its impact is achieved at low deposited fluences, as shown for $\lambda_{\mathrm{c}}=800\,\si{\nano\meter}$. Enhanced absorption and short pulses are beneficial. Fluences above the optimum switching point diminish the switching process through thermal fluctuations. However, the growing contribution of $\Delta M^{\mathrm{ind}}$ compensates for the declined contribution of the MCD at higher electron temperatures over a wide temperature range. This interaction creates a narrow fluence window of $F_{\mathrm{abs}}\approx 0.5\,\si{\milli\joule\per\square{\centi\meter}}$ between the threshold and maximum switching points within which both processes operate cooperatively enhancing the process. Generally, the maximum of $\Delta M/M_{\mathrm{S}}$ emerges when $\Delta M^{\mathrm{ind}}\approx 0.4 - 0.6\,\si{\micro_B}$. Simultaneously, Simultaneously, $\Delta M/M_{\mathrm{S}}$ reaches a larger value for lower deposited energy density. In reference \cite{Shalaby.2018}, the authors discuss a similar competition between induced magnetization and demagnetization for the THz spectral range. There, coherent excitations generated at low pump fluences enable spin manipulation. At higher fluences, additional demagnetization suppresses the process.

\begin{acknowledgments}
We are grateful to the German Science Foundation (DFG) for financial support through the program: 'Fundamental aspects of all-optical single pulse switching in nanometer-sized magnetic storage media' Project number: 439225584.
E.S. was supported by TERAFIT project No. $\mathrm{CZ}.02.01.01/00/22\textunderscore008/0004594$ funded by OP JAK, call Excellent Research. 
K.C. supported by the Czech Science Foundation grant no. 23-­04746S.
Further, we thank Tiffany Santos for the granular FePt hard disk media.
\end{acknowledgments}

\section*{Data Availability Statement}

The data that support the findings of
this study are available from the
corresponding author upon reasonable
request.

%\nocite{*}
\bibliography{sources}% Produces the bibliography via BibTeX.
	
\newpage
\renewcommand{\thefigure}{S\arabic{figure}} 

\section*{Supplementary material}  

\begin{figure}[h]
	\centering
	\includegraphics[width=0.8\linewidth]{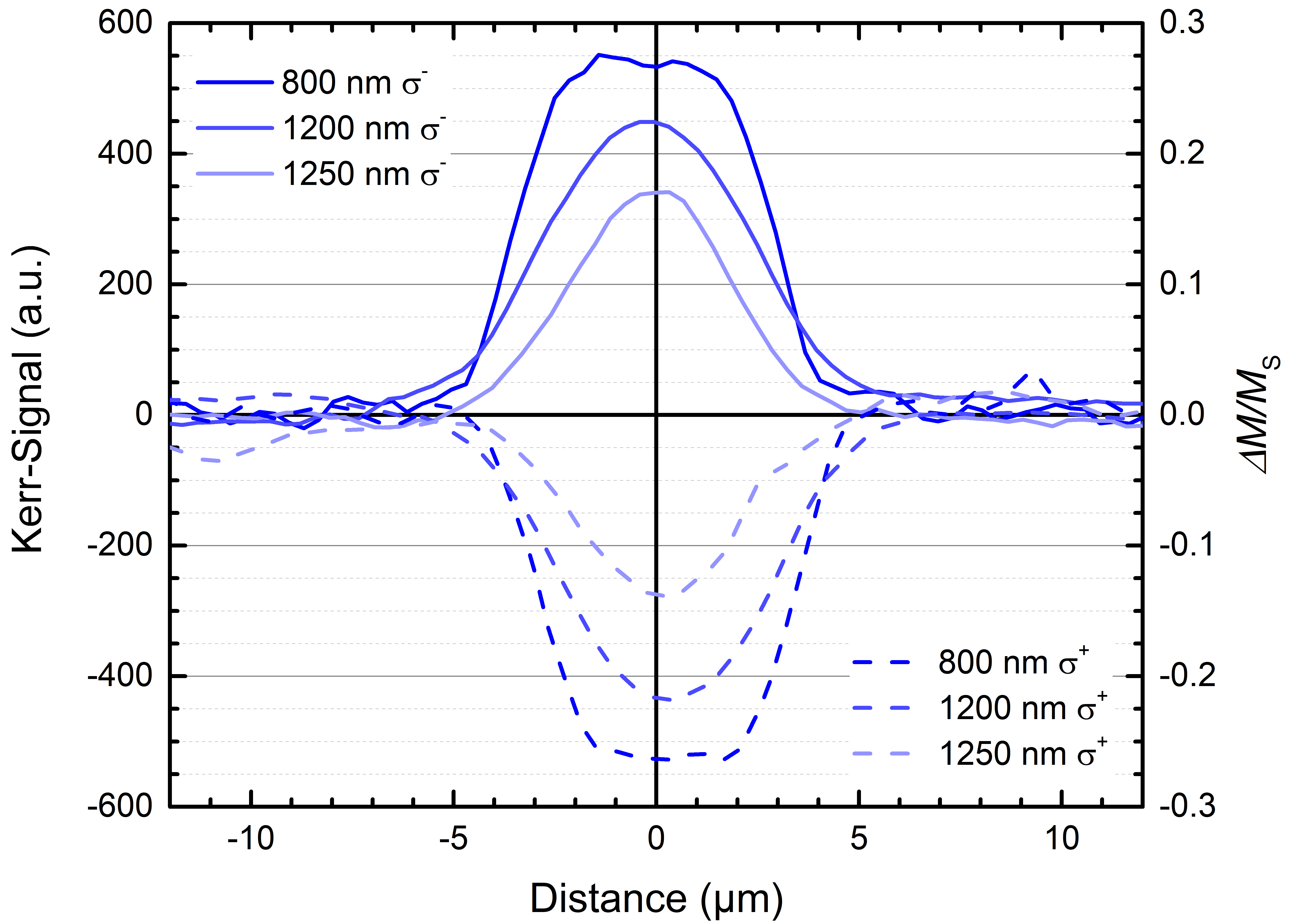}
	\caption{Symmetric switching behavior $\lambda_{\mathrm{c}}=800\,\si{\nano\meter}-1250\,\si{\nano\meter}$ for both helicities denoted by the solid lines ($\sigma^-$) and dashed lines ($\sigma^+$).}
	\label{fig:symswitch1}
\end{figure}

\begin{figure}%[H]
	\centering
	\includegraphics[width=0.8\linewidth]{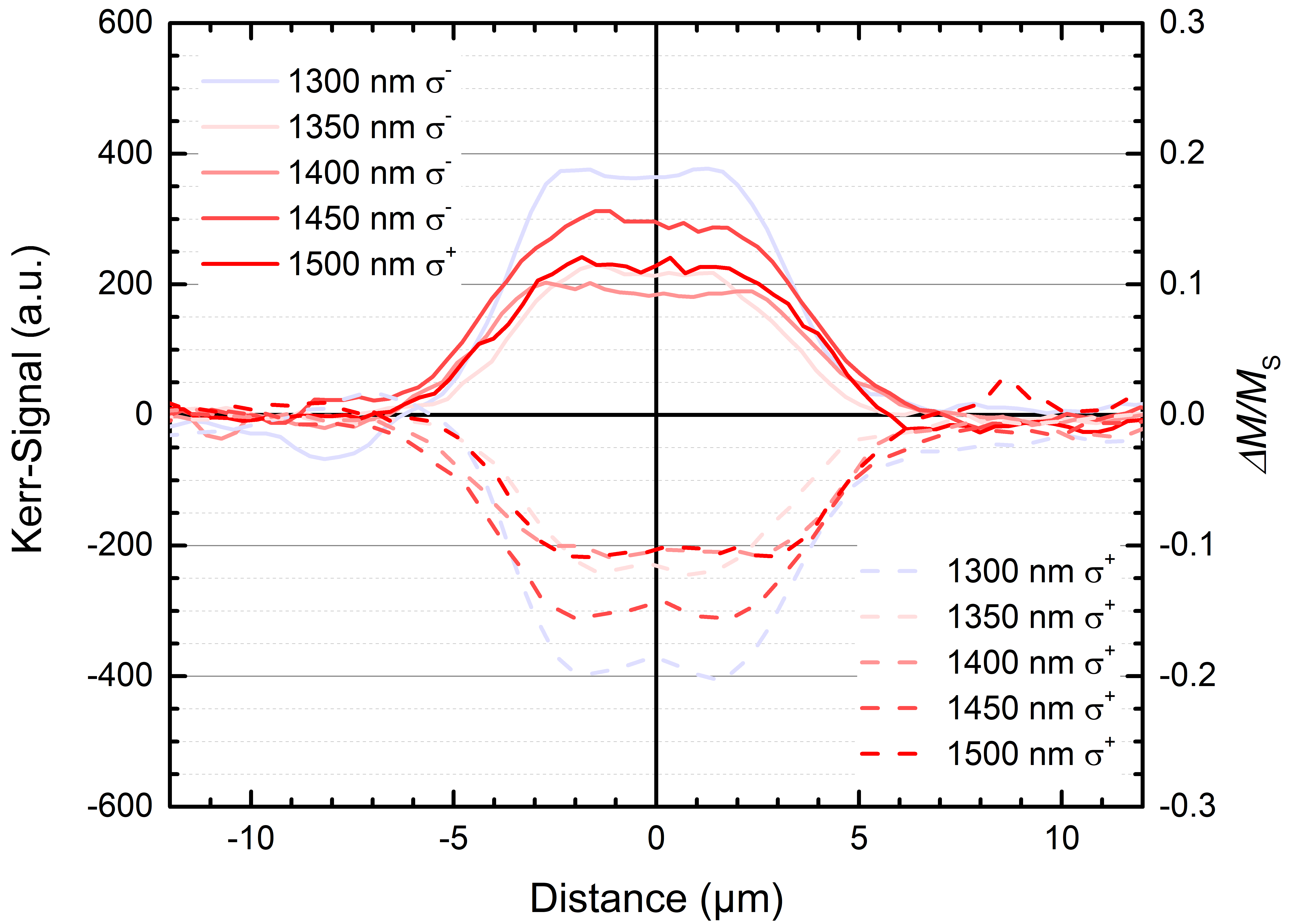}
	\caption{Symmetric switching behavior for  $\lambda_{\mathrm{c}}=1300\,\si{\nano\meter}-1500\,\si{\nano\meter}$ for both helicities denoted by the solid lines ($\sigma^-$) and dashed lines ($\sigma^+$).}
	\label{fig:symswitch2}
\end{figure}

\begin{figure}%[H]
	\centering
	\includegraphics[width=0.7\linewidth]{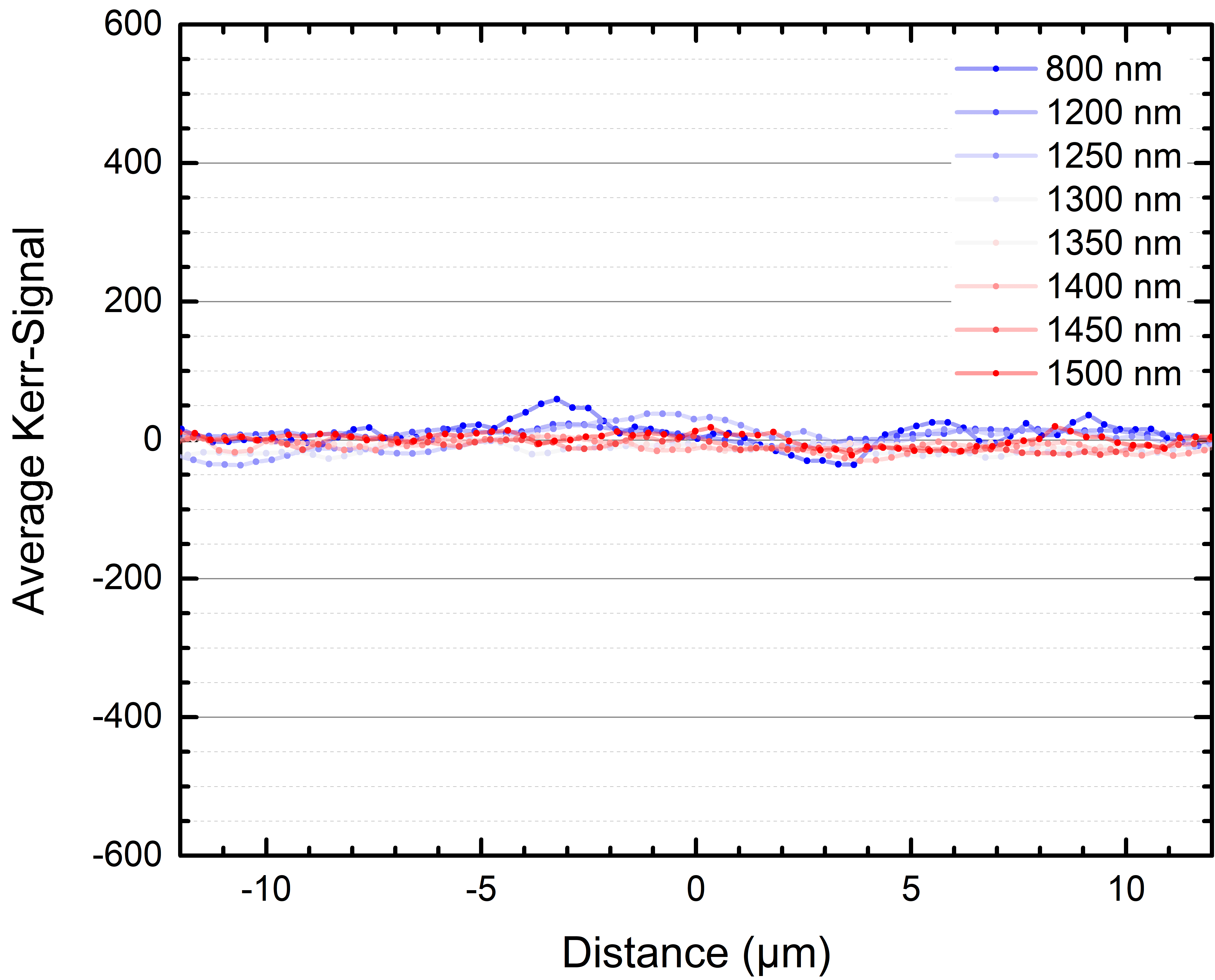}
	\caption{Symmetric switching behavior, the difference of both switching magnitudes remains within the background noise of the recorded Kerr images.}
	\label{fig:symswitchaverage}
\end{figure}

Figs. \ref*{fig:symswitch1} and \ref*{fig:symswitch2} show the symmetric switching behavior for both helicities and central wavelengths $\lambda_{\mathrm{c}}=800\,\si{\nano\meter}-1250\,\si{\nano\meter}$ and $\lambda_{\mathrm{c}}=1300\,\si{\nano\meter}-1500\,\si{\nano\meter}$, respectively. All profiles stem from maximum switching rate fluence experiments for each wavelength, starting with a demagnetized sample. For the shorter wavelengths in Fig. \ref*{fig:symswitch1}, the maximum switching rate $\Delta M/M_{\mathrm{S}}$ decreases gradually, and the profile shape is close to the Gaussian distribution with a maximum in the center. This shape mirrors the intensity of the Gaussian beam profile. For the longer wavelengths in Fig. \ref{fig:symswitch2}, the maximum $\Delta M/M_{\mathrm{S}}$ values and their decline with increasing wavelength are smaller and less distinctive. The profile shapes change, and the maximum in the center broadens to a plateau. The switching behavior remains symmetric, and the magnitude of $\Delta M/M_{\mathrm{S}}$ for each helicity, but the direction is opposite. Fig. \ref*{fig:symswitchaverage} confirms that the difference between the switching magnitudes or the average signal remains within the background noise.

\end{document}